\documentclass[aps,prb,twocolumn,showpacs]{revtex4}

\usepackage{hyperref}
\usepackage{graphicx}

\renewcommand{\vec}[1]{{\mathbf{#1}}}
\newcommand{\un}[1]{{\,\mathrm{#1}}}

\begin{document}

\title{Polarization analysis of K-edge resonant x-ray scattering of
  germanium}

\author{C. Detlefs} 

\affiliation{European Synchrotron Radiation Facility, Bo{\^\i}te
  Postale 220, 38043 Grenoble, Cedex, France.}

\date{\today}

\begin{abstract}
  The polarization of K-edge resonant scattering at the space group
  ``forbidden'' $(0~0~6)$ reflection of $\mathrm{Ge}$ was measured as
  function of the azimuthal angle, $\psi$. The experimental results
  are compared to model calculations based on symmetry analysis of the
  resonant scattering tensors.
\end{abstract}

\maketitle

In the last few years the investigation of resonant scattering
phenomena has allowed novel studies of antiferromagnetism (through
resonant magnetic x-ray scattering\cite{Gibbs88}) and, more recently,
of orbital \cite{Murakami98,Paolasini99,McMorrow01,Paixao02} and
quadrupolar order \cite{Hirota00}. Although much progress has been
made in the theory and interpretation of these effects, a large number
of open questions concerning the origin of the scattering remains.

A characteristic feature of resonant scattering is its polarization
dependence, which differs markedly from that of normal Thompson
scattering. Indeed, polarization analysis (PA) has now been developed
into a standard tool for the study of the aforementioned effects.  In
this communication I present a study of resonant x-ray scattering at
the $\mathrm{Ge}$ K-edge with emphasis on the analysis of polarization
effects.  Anisotropic tensor susceptibility (ATS) scattering
\cite{Templeton85,Dmitrienko83,Dmitrienko84} and orbital/quadrupolar
order resonances are intimately related as the scattering arises from
the same transitions between core level and valence band electronic
states \cite{Blume94}. Therefore systematic studies of a well-known
reference system may lead to better understanding of the complex
situation in compounds exhibiting orbital order.

One may distinguish three different classes of ATS scattering,
depending on the rank of the scattering tensor: The original
experiment on $\mathrm{NaBrO_3}$ may be described by second rank
tensors corresponding to electric dipole (E1) transitions
\cite{Templeton85,Dmitrienko83,Dmitrienko84}.  A later experiment on
$\alpha-\mathrm{Fe_2O_3}$ evidenced electric quadrupole (E2)
transitions, which give rise to fourth rank tensors
\cite{Finkelstein92}.  Finally, ATS scattering in $\mathrm{Ge}$ was
attributed to rank three tensors.  Two different origins of this
tensor were proposed: An E1--E2 mixed
resonance\cite{Templeton94,Elfimov02,Elfimov02b}, and an E1--E1
process combined with a displacement of the scattering atom due to
thermal motion\cite{Kokubun01,Kirfel02}. The subject is still under
discussion\cite{Dmitrienko03,Elfimov03}.

The azimuthal dependence of the scattered beam intensity was
calculated and experimentally verified by \citeauthor{Templeton94}
\cite{Templeton94}, but without polarization analysis. For some
selected photon energies, the phase of the $(0~0~6)$ and $(2~2~2)$
resonant scattering was determined by \citeauthor{Lee01}
\cite{Lee01} through the interference with Umweg reflections. Finally,
\citeauthor{Kokubun01}\cite{Kokubun01} and
\citeauthor{Kirfel02}\cite{Kirfel02} studied the temperature
dependence of the $(0~0~2)$ and $(0~0~6)$ resonant scattering. They
observed a strong increase of the intensity with increasing
temperature, but only minor variations in the line shape of the
resonance. They concluded that the dominant origin of resonant
scattering lies in anisotropic thermal motion of the $\mathrm{Ge}$
atoms.

The aim of the present experiment was to complement the existing body
of experimental data to further study resonant x-ray scattering of
odd-rank tensors. Presented below are measurements of the resonant
line shape and the azimuthal dependence of the polarization of the
$(0~0~6)$ reflection, which is forbidden by the glide-plane extinction
rule.

The experiment was performed at the magnetic scattering beamline,
ID20, of the ESRF\@. The scattering geometry was vertical, with
incident $\sigma$ polarization. A single crystal of $\mathrm{Ge}$ with
a polished $(0~0~L)$ surface normal was mounted in the azimuthal scan
configuration which allows to turn the sample about the scattering
vector, $\vec{Q}$. The diffracted beam was reflected by a $\mathrm{Au}
(3~3~3)$ polarization analyzer (PA) with $2\theta_{\mathrm{PA}}
\approx 90^\circ$, which also rejected fluorescence and other diffuse
background. The PA may be rotated about the diffracted beam (angle
$\eta$). $\eta=0$ when the diffraction planes of the PA and the sample
coincided, i.e.{\ }when the PA accepted $\sigma^\prime$ polarization.

\begin{figure}[tp]
  \centerline{%
  \includegraphics[width=0.81\columnwidth,height=0.81\textheight,keepaspectratio=true]{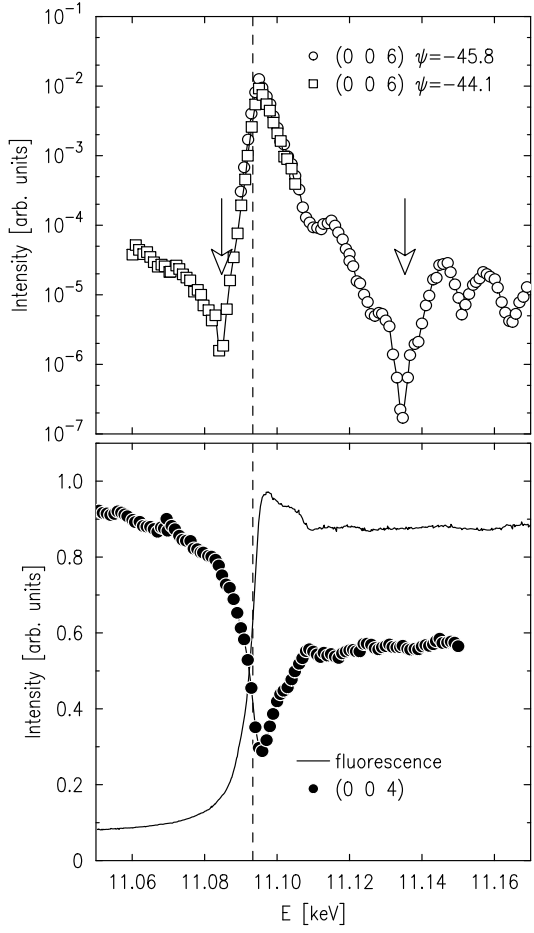}
  } \caption[]{\label{Escan}Energy scans through the $\mathrm{Ge}$ K
  absorption edge, with the PA set at $\eta=0$. The scattering at the
  forbidden $(0~0~6)$ reflection shows a sharp resonance (top) near
  the inflection point of the absorption, as determined from
  measurements of the fluorescence and the $(0~0~4)$ Bragg peak
  (bottom). The dashed line indicates the inflection point of the
  fluorescence curve. }
\end{figure}

Fig.~\ref{Escan} (top) shows the intensity of the $(0~0~6)$ reflection
as function of the incident photon energy. For each energy, the
intensity was determined by integrating a rocking scan after
subtraction of a constant background. To avoid contamination by strong
Umweg reflections the energy dependence was measured in several scans
at different azimuthal angles. Regions where a marked azimuthal
dependence on was observed were rejected. No absorption correction was
applied.  The fluorescence yield and the intensity of the strong,
allowed $(0~0~4)$ reflection are shown in Fig.~\ref{Escan}(bottom) for
comparison.

Several features of these data are worth further discussion:
Significant intensity is observed below the edge. This scattering may
be due to the tails of Umweg reflections\cite{Tischler86}, or to
non-resonant ATS scattering\cite{Amara98,Amara01}. Furthermore, two
deep minima (indicated by the arrows in Fig.~\ref{Escan}), below and
above the main resonance are observed. Such minima are characteristic
of a change of sign of the scattering amplitude and might indicate
interference between resonant and non-resonant contributions. It would
therefore be of interest to determine the phase of the resonant
scattering, e.g.{\ }through interference with an Umweg reflection of
known phase\cite{Tischler86,Lee01}. Finally, towards higher energies
oscillations reminiscent of DAFS (Diffraction Anomalous Fine
Structure) set in.

For the polarization measurements discussed in the remainder of this
Communication the photon energy was tuned to the maximum of the
resonance at $E=11.096\un{keV}$.

\begin{figure}[tp]
  \centerline{%
  \includegraphics[width=0.81\columnwidth,height=0.76\textheight,
  keepaspectratio=true]{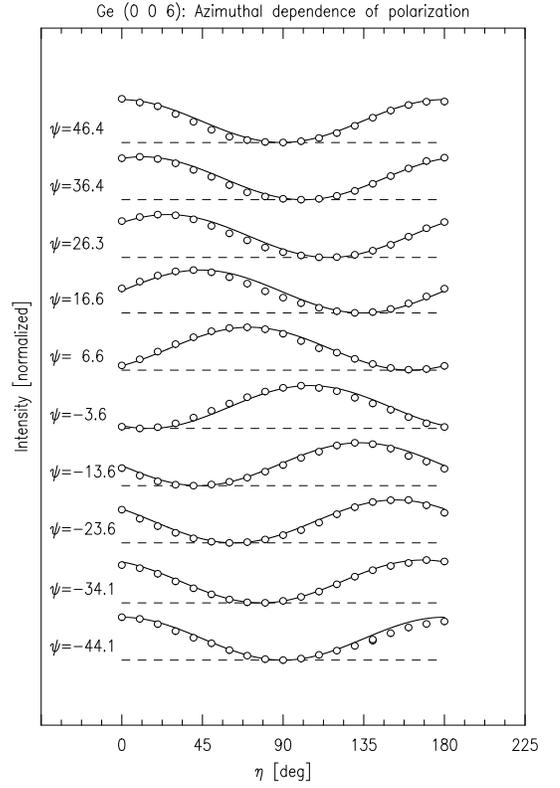} }
  \caption[]{\label{cascade}Integrated intensities as function of the
  PA angle, $\eta$, for different azimuthal angles, $\psi$. All data
  are taken at $E=11.096\un{keV}$. Each open circle represents the
  intensity obtained by integrated a rocking scan of the analyzer
  crystal. The solid lines represent model calculation based in
  eqs.~\ref{p1theo},~\ref{p2theo}, and~\ref{pa}.}
\end{figure}

Fig.~\ref{cascade} shows the detected intensity as a function of the
orientation of the PA for different azimuthal angles, $\psi$. For each
$\psi$, the $\mathrm{Ge} (0~0~6)$ reflection was carefully aligned,
and it was verified that there was no Umweg excitation in the
immediate vicinity. This requirement resulted in slightly irregular
values of $\psi$. The sample was then kept in this position while the
polarization of the scattered beam was measured by rocking the
analyzer crystal for settings of $\eta$ between $0^\circ$ and
$180^\circ$. The integrated intensities of these rocking scans are
shown as open circles in Fig.~\ref{cascade}. The data for each $\psi$
are normalized so that they are independent of the intensity of the
scattered beam. Reliable measurements of the intensity of the
scattered beam proved difficult in the present experimental setup, as
the resolution function is determined by the combined narrow angular
acceptances of the sample and the analyzer.

The polarization of an x-ray beam is most conveniently described by
the Stokes parameters \cite{Lipps54},
\begin{eqnarray}
  P_1
  & = & 
  \frac{
    \left|F_{\sigma\sigma^\prime}\right|^2 
    - 
    \left|F_{\sigma\pi^\prime}\right|^2
    }{
    \left|F_{\sigma\sigma^\prime}\right|^2 
    + 
    \left|F_{\sigma\pi^\prime}\right|^2
    }
  \label{p1}
  \\
  P_2
  & = & 
  \frac{
    \left | F_{\sigma\sigma^\prime} + F_{\sigma\pi^\prime} \right|^2
    - \left | F_{\sigma\sigma^\prime} - F_{\sigma\pi^\prime} \right|^2
    }{
    2\left(
      \left|F_{\sigma\sigma^\prime}\right|^2 
      + 
      \left|F_{\sigma\pi^\prime}\right|^2
    \right)    
    }
  \label{p2}
  \\
  P_3
  & = & 
  \frac{
    \left | F_{\sigma\sigma^\prime} + i F_{\sigma\pi^\prime} \right|^2
    - \left | F_{\sigma\sigma^\prime} - i F_{\sigma\pi^\prime} \right|^2
    }{
    2\left(
      \left|F_{\sigma\sigma^\prime}\right|^2 
      + 
      \left|F_{\sigma\pi^\prime}\right|^2
    \right)    
    }
\end{eqnarray}
The dependence of the transmission of an idealized linear PA on these
is given by
\begin{equation}
  I(\eta) 
  \propto
  \left[1 + P_1 \cos(2\eta) + P_2 \sin(2\eta) \right]
  \label{pa}
\end{equation}

\begin{figure}[tp]
  \centerline{%
    \includegraphics[width=0.81\columnwidth]{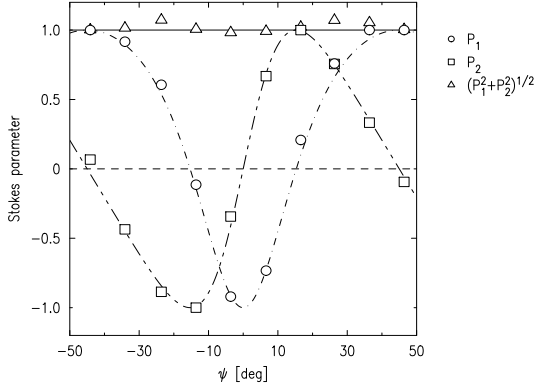}
    }
  \caption[]{\label{stokes}Stokes parameters, $P_{1,2}$
    determined from fits of the intensities to eq.~\ref{pa}, compared
    to the model, eqs.~\ref{p1theo} and~\ref{p2theo}.}
\end{figure}

The data presented in Fig.~\ref{cascade} where fitted to eq.~\ref{pa}
in order to determine the dependence of the Stokes parameters, $P_1$
and $P_2$, on $\psi$. The results are shown in Fig.~\ref{stokes}.
While the degree of circular polarization, $P_3$, was not directly
measured, it vanishes because $\left|P_3\right| \leq 1 -
\sqrt{P_1^2+P_2^2} \approx 0$. 

As pointed out above, the dependence of the polarization on the
azimuthal angle may be calculated from symmetry principles alone.

In $\mathrm{Ge}$ the two sites within the primitive unit cell are
related by an inversion center at the midpoint of the covalent
bond. As even rank tensors are invariant under space inversion, the
aforementioned E1 and E2 resonances do not contribute to scattering at
``forbidden'' reflections, $(H K L)$ with $H+K+L=4n+2$. An odd rank
tensor is therefore needed to explain resonant scattering observed at
forbidden reflections \cite{Templeton94,Elfimov02,Elfimov02b}.

The E1--E2 process\cite{Templeton94,Elfimov02,Elfimov02b} directly
gives rise to third rank tensors, $A_{\alpha\beta\gamma}$. The
resulting atomic scattering amplitude may be written as \cite{Blume94}

\begin{equation}
  F(\vec{k},\vec{\epsilon},\vec{k^\prime},\vec{\epsilon^\prime}) 
  \propto
  \epsilon^{\prime\star}_\alpha 
  \epsilon_\beta 
  \left[
    A_{\alpha\beta\gamma} k_\gamma
    -
    A^\prime_{\alpha\beta\gamma} k^\prime_\gamma
  \right]
  \label{amp}
\end{equation}
with
\begin{eqnarray}
  A_{\alpha\beta\gamma}
  & = &
  \sum_{a,b}
  B^{(a,b)}(\hbar\omega)
  C^{(a,b)}_{\alpha\beta\gamma}
  \\
  A^\prime_{\alpha\beta\gamma}
  & = &
  \sum_{a,b}
  B^{(a,b)}(\hbar\omega)
  C^{(a,b)\star}_{\beta\alpha\gamma}
  \\
  B^{(a,b)}(\hbar\omega)
  & = &
  \frac{p_a}{E_a-E_b+\hbar\omega+i\Gamma/2}
  \\
  C^{(a,b)}_{\alpha\beta\gamma}
  & = &
  \left< a \left|
      r_\alpha
    \right| b \right>
  \left< b \left|
      r_\beta r_\gamma
    \right| a \right>,
  \label{defC}
\end{eqnarray}
where $\epsilon$ ($\epsilon^\prime$) and $k$ ($k^\prime$) are the
polarization and wave vectors of the incident (scattered) beam.
$\left|a\right>$ and $\left|b\right>$ are the initial (=final) and
intermediate electronic states. $\hbar \omega$ is the photon energy,
and $\Gamma$ the inverse life time of the excited state. $p_a$ is the
probability that the corresponding state is occupied.

In a system that is invariant under time reversal states which are
related by time reversal, $\left|\bar{a}\right>=T\left|a\right>$, have
the same energy, $E_{\bar{a}}=E_a$, and probability of being occupied,
$p_{\bar{a}}=p_a$, so that $B^{(\bar{a},\bar{b})} = B^{(a,b)}$.
Furthermore, $\left<\bar{a}\left| r_\alpha \right|\bar{b}\right> =
\left<b\left| r_\alpha \right|a\right>$ and
$\left<\bar{a}\left|r_\beta r_\gamma \right|\bar{b}\right> =
\left<b\left| r_\beta r_\gamma \right|a\right>$, so that
$C^{(\bar{a},\bar{b})}_{\alpha\beta\gamma} =
C^{(a,b)\star}_{\alpha\beta\gamma}$. The sum over $a$ and $b$ may
equally well be carried out over $\bar{a}$ and $\bar{b}$, therefore
\cite{Blume94}
\begin{eqnarray}
  A_{\alpha\beta\gamma}
  & = &
  \frac{1}{2}
  \sum_{a,b}
  B^{(a,b)}(\hbar\omega)
  \left[
    C^{(a,b)}_{\alpha\beta\gamma}
    +
    C^{(\bar{a},\bar{b})}_{\alpha\beta\gamma}
  \right]
  \\
  & = &
  \sum_{a,b}
  B^{(a,b)}(\hbar\omega)
  \Re(C^{(a,b)}_{\alpha\beta\gamma})
  \\
  A^\prime_{\alpha\beta\gamma}
  & = &
  A_{\beta\alpha\gamma},
\end{eqnarray}
where $\Re(x)$ denotes the real part of $x$. With this, eq.~\ref{amp}
reduces to
\begin{eqnarray}
  F(\vec{k},\vec{\epsilon},\vec{k^\prime},\vec{\epsilon^\prime}) 
  & \propto &
  \epsilon^{\prime\star}_\alpha 
  \epsilon_\beta 
  \left[
    A_{\alpha\beta\gamma} k_\gamma
    -
    A_{\beta\alpha\gamma} k^\prime_\gamma
  \right]
  \\
  & = &
  \epsilon^{\prime\star}_\alpha 
  \epsilon_\beta 
  \left[
    -\left(
      A_{\alpha\beta\gamma} + A_{\beta\alpha\gamma}
    \right) Q_\gamma
    \right. \nonumber \\ & & \left.
    +
    \left(
      A_{\alpha\beta\gamma} - A_{\beta\alpha\gamma}
    \right)
    \left(
      k_\gamma + k^\prime_\gamma
    \right)
  \right],
  \label{time}
\end{eqnarray}
where $\vec{Q}=(H K L)=\vec{k}^\prime-\vec{k}$ is the scattering
vector. The first term was already discussed by
\citeauthor{Templeton94} \cite{Templeton94}. The second term does not
necessarily vanish in all cases --- in fact, for certain symmetries it
may lead to x-ray natural circular dichroism, XNCD \cite{Goulon98}.

A different explanation for the origin of resonant scattering in
$\mathrm{Ge}$, termed Thermal Motion Induced (TMI) scattering, was
recently proposed by
\citeauthor{Dmitrienko00}\cite{Dmitrienko00}. Their theory constructs
a rank 3 tensor from E1 transitions combined with a displacement of
the scattering atom out of the crystallographic, high symmetry
position.  These displacements are assumed to be of thermal origin, so
that the scattered intensity increases strongly with rising
temperature. This increase has indeed been observed experimentally
\cite{Kokubun01,Kirfel02}.

In K-edge resonances an electron is promoted from a $1s_{1/2}$ core
level into an \emph{unoccupied} valence band. The E1 and E2 selection
rules require $\Delta l=1$ and $\Delta l=2$, respectively.
Consequently, the E1--E2 mixed and TMI resonances differ in their
sensitivity to the the conduction band symmetry: The former probes
only valence bands with contributions of both $p$ and $d$ character,
while the latter requires $p$ character, only.  In both cases band
structure calculations are needed to obtain the spectral shape of the
resonance \cite{Elfimov02,Elfimov02b}.

However, as pointed out above, detailed knowledge of the matrix
elements is not necessary to calculate the polarization properties of
the resonances.  It suffices to require that $A_{\alpha\beta\gamma}$
is invariant under the point group of the scattering site, $T_d$ for
the case of $\mathrm{Ge}$. The resulting tensor is
$A_{\alpha\beta\gamma} = D T_{\alpha\beta\gamma}$, where $D$ is a
complex number depending of the matrix elements, the resonant
denominators, the densities of state, etc, but not on the scattering
geometry, and the photon polarization and wave vectors.
$T_{\alpha\beta\gamma}$ is symmetric over all its indices, i.e.,
$T_{xyz}=T_{xzy}=T_{yzx}=T_{yxz}=T_{zxy}=T_{zyx}=1$ and $0$ otherwise
\cite{Kirfel02}.  In particular, $A_{\alpha\beta\gamma}$ is symmetric
in $\alpha$ and $\beta$, so that the second term in eq.~\ref{time}
vanishes, whereas the first term gives
\begin{equation}
  F 
  \propto 
  D\,
  \vec{\epsilon}^{\prime\star}
  \cdot
  \left(\begin{array}{ccc}
      0 & L & K \\
      L & 0 & H \\
      K & H & 0 
    \end{array}\right)
  \cdot \vec{\epsilon}.
  \label{F}
\end{equation}

Eq.~\ref{F} describes the third-rank tensor resonant contribution to
any Bragg reflection. In general, this contribution is much smaller
than the Thompson scattering. For practical purposes this term is
therefore significant only when the Thompson contribution vanishes,
i.e., at reflections which are ``forbidden'' due to glide plane or
screw axis extinction rules, or structure factor
arithmetic\cite{Templeton94}.

For $(0~0~L)$-type reflections of $\mathrm{Ge}$, the anomalous
resonant scattering amplitude is proportional to
\begin{equation}
  F
  \propto
  D\, Q
  \left(
    \epsilon^{\prime\dagger}_1
    \epsilon_2
    +
    \epsilon^{\prime\dagger}_2
    \epsilon_1
  \right),
\end{equation}
where the polarization vectors have to be transformed into the
coordinate system of the crystal, i.e., the dependence on $\psi$ is
implicit in $\vec{\epsilon}$ and $\vec{\epsilon}^\prime$.  

The polarization of the scattered beam is completely described by the
scattering amplitudes into the channels with polarization
perpendicular ($\sigma^\prime$) and parallel ($\pi^\prime$) to the
scattering plane. For incident $\sigma$ polarization they are
\begin{eqnarray}
  F_{\sigma\sigma^\prime}(\theta,\psi) 
  & = & 
  D\, 
  Q \sin(2\psi)
  \label{fsigma}
\\
  F_{\sigma\pi^\prime}(\theta,\psi) 
  & = & 
  D\,
  Q \sin(\theta)\cos(2\psi),
  \label{fpi}
\end{eqnarray}
with $\psi=0$ when the azimuthal reference vector, chosen as
$\vec{h}_0=(1~0~0)$, lies within the scattering plane. 

Equations~\ref{fsigma} and~\ref{fpi} yield the Stokes parameters
\begin{eqnarray}
  P_1(\theta,\psi)
  & = & 
  \frac{
    \sin^2(2\psi) - \sin^2(\theta)\cos^2(2\psi)
    }{
    \sin^2(2\psi) + \sin^2(\theta)\cos^2(2\psi)
    }
  \label{p1theo}
  \\
  P_2(\theta,\psi)
  & = &
  \frac{
    \sin(4\psi)\sin(\theta)    
    }{
    \sin^2(2\psi) + \sin^2(\theta)\cos^2(2\psi)
    }
  \label{p2theo}
  \\
  P_3(\theta,\psi)
  & = &
  0.
\end{eqnarray}
This result is identical to that of \citeauthor{Elfimov02b}
\cite{Elfimov02b}, obtained from band structure calculations.  Note
that $P_{1,2,3}$ are independent of the scaling factor, $D$, and that
$P_1^2+P_2^2=1$ for all $\psi$. The scattered beam is therefore always
linearly polarized. The experimental data, shown in Fig.~\ref{stokes},
agree well with the values calculated from eqs.~\ref{p1theo}
and~\ref{p2theo} --- bearing in mind that there are no adjustable
parameters.

In summary, I have presented measurements of the polarization of ATS
scattering at the forbidden $(0~0~6)$ reflection of $\mathrm{Ge}$. The
dependence of the Stokes parameters $P_1$ and $P_2$ on the azimuthal
angle, $\psi$, is well described by the model of third-rank scattering
tensors. The line shape of the resonance was measured over an extended
energy range.

In high symmetry systems such as this one, symmetry analysis of the
scattering tensor allows detailed predictions about variation of the
scattered intensity and polarization with the azimuthal angle. The
technique is therefore particularly well adapted to the study of
orbitally ordered systems, where the determination of the symmetry of
the order parameter is a problem of fundamental importance.

\acknowledgments

Discussions with E. Lorenzo-Diaz, G. Sawatzky and L. Paolasini are
gratefully acknowledged. Furthermore, I thank K. D. Finkelstein for
critical reading of the manuscript and helpful comments.


\end{document}